\documentclass[10pt, a4paper, twocolumn]{article}

\usepackage[utf8]{inputenc}    % Encoding
\usepackage[T1]{fontenc}       % Font encoding
\usepackage{lmodern}           % Latin Modern font
\usepackage[left=1.5cm, right=1.5cm, top=1.785cm, bottom=2.0cm]{geometry}

\usepackage{setspace}          % Line spacing
\usepackage{authblk}           % Authors and affiliations
\usepackage{graphicx}          % Figures
\usepackage{amsmath, amssymb}  % Math
\usepackage{hyperref}          % Hyperlinks
\usepackage{abstract}          % Abstract formatting
\usepackage{titlesec}          % Section title formatting (optional)
\usepackage[numbers,sort&compress]{natbib}
\usepackage{balance}
\usepackage{lscape}

\hypersetup{
    colorlinks=true,
    linkcolor=blue,
    citecolor=blue,
    urlcolor=blue
}

\title{\textbf{Microfluidic platform for biomimetic tissue design and multiscale rheological characterization}}

\author[1,$^{\ddag}$]{Majid Layachi}
\author[1]{Remi Merindol}
\author[1]{Laura Casanellas$^{*}$}
\affil[1]{Laboratoire Charles Coulomb, Place Eug\`{e}ne Bataillon,34095  Montpellier, France}
\affil[$^{\ddag}$]{Current affiliation: ULB-TIPs, C.P. 165/67,  Avenue F.D. Roosevelt, 50,  1050 Brussels, Belgium}
\affil[$^{*}$]{Email: laura.casanellas-vilageliu@umontpellier.fr}

\date{}

\begin{document}

% --- Title and Abstract in One Column ---
\twocolumn[
\begin{@twocolumnfalse}
    \maketitle

    \begin{abstract}
    The way living tissues respond to external mechanical forces is crucial in physiological processes like embryogenesis, homeostasis or tumor growth. Providing a complete description across length scales which relates the properties of individual cells to the rheological behavior of complex 3D-tissues remains an open challenge. 
The development of simplified biomimetic tissues capable of reproducing essential mechanical features of living tissues can help achieving this major goal. We report in this work the development of a microfluidic device that enables to achieve the sequential  assembly of biomimetic prototissues and their rheological characterization. 
We synthesize prototissues by the controlled assembly of Giant Unilamellar Vesicles (GUVs) for which we can tailor their sizes and shapes as well as their level of GUV-GUV adhesion.
We address a rheological description at multiple scales which comprises an analysis at the local scale of individual GUVs and at the global scale of the prototissue. 
The flow behavior of prototissues ranges from purely viscous to viscoelastic for increasing levels of adhesion. 
At low adhesion the flow response is dominated by viscous dissipation, which is mediated by GUV spatial reorganizations at the local scale, whereas at high adhesion the flow is viscoelastic, which results from a combination of internal reorganizations and deformation of individual GUVs. 
Such multiscale characterization of model biomimetic tissues provides a robust framework to rationalize the role of cell adhesion in the flow dynamics of living tissues.
    \end{abstract}

    \vspace{1em} % Adds a small space before the main text starts
\end{@twocolumnfalse}
]

%%%MAIN TEXT%%%%
\section{Introduction}
\label{intro}
The rheological properties of living tissues determine the way they respond to external mechanical stimuli. 
There is experimental and theoretical evidence that rheological features are relevant for the correct achievement of physiological processes in which collective spatiotemporal reorganizations of cells are involved, such as embryogenesis or tumor growth \cite{Bershadsky2003,Jain2014,Mongera2018,Petridou2019,Corominas-Murtra2021}. 
In these complex processes there exists an intricate relationship  between the rheological properties of the overall tissue with the mechanical properties of the individual constituting cells and their adhesiveness \cite{Maitre2013,Streichan2018,Kashkooli2021}. %
Establishing the link across scales between the rheological behavior of living tissues (and more generally of soft materials) at the macroscopic scale with their constituents at a mesoscopic scale still represents a major scientific challenge \cite{Verdier2003,Stooke-Vaughan2018,Collinet2021}.  
In living tissues this relationship across length-scales is complexified by feedback mechanisms, by which cells can sense external mechanical stresses and remodel their cytoskeletal network or the adhesion affinity with their neighboring cells or with the extracellular matrix \cite{Bershadsky2003}. 
The development of biomimetic simplified tissues, devoid of any active processes but which are capable of reproducing essential mechanical properties of living tissues, facilitates probing the role of specific physical mechanisms and provides a quantitative framework enabling to test them systematically and independently. 
A diversity of artificial model tissues have been designed over the last decade, based on the controlled assembly of biomimetic cell models into higher order spatial arrangements.
The architecture of biomimetic prototissues can range from  minimalistic models mainly driven by surface tension, such as soap foams \cite{Dollet2006,Dollet2010} or oil-in-water emulsions \cite{Golovkova2021}, up to \textit{in-vitro} spheroids constituted of living cells \cite{Lin2008,Marmottant2009,Alessandri2013} or even more sophisticated organoids capable of reproducing targeted functionalities of specific organs \cite{Ahadian2017,Zhao2022}.   
The flow of faceted soft materials (from soap foams to living tissues) is mediated by the occurrence of rearrangements, at the microscopic level, between the constituting subunits (bubbles, droplets or cells) \cite{Chen2012,Cohen-Addad2013,Dollet2014}. This comes along with the elastic deformation of the subunits themselves. Experiments with emulsions at the jammed state enabled to reveal the intricate relationship that exists between the spatial localization of rearrangements (mainly T1 events) and the deformation field of the droplets \cite{Desmond2015}. 
The addition of adhesion between droplets was shown, in turn, to impact the dynamics and topology of rearrangements \cite{Montel2022,Golovkova2020,Golovkova2021}. 
Likewise, in developing living tissues there exists a complex relationship between cell deformation and cell-cell rearrangements which can be modulated via spatial variations of cell adhesion, and which is crucial for morphogenetic processes \cite{Butler2009,Xie2018}. 

In our team, we recently developed and characterized a tissue prototype making use of Giant Unilamellar Vesicles (GUVs) as constituting units (that we name hereafter \textit{vesicle prototissues} \cite{Casas-Ferrer2021,Layachi2022}).  
GUVs are constituted of a lipid bilayer which can reproduce the essential mechanical properties of cell membranes \cite{Fenz2012}. 
However, vesicle are passive systems which cannot display any active mechanisms such as mechano-sensing or protrusion formation. 
Adhesion between GUVs can be controlled using specific bonding in order to provide cohesive prototissues. 
In particular, the specific recognition between streptavidin and biotin molecules has been largely used in the literature, due to its robustness and high affinity \cite{Chiruvolu1994,Kisak2000,Amjad2017b}, as well as the use of complementary DNA strands, due to the larger degrees of tunability offered by this technology \cite{Parolini2015a,Beales2007a,Hadorn2010a,Liu2023,Huang2024}. 
In our previous work we used a simple bulk assembly protocol for which we adjusted the mixing method and the concentration of GUVs and adhesion molecules \cite{Casas-Ferrer2021}.  
Synthesized prototissues using this protocol displayed spheroidal or sheet-like morphologies and controlled typical sizes and degrees of adhesion between GUVs. 
However, bulk self-assembly provided vesicle prototissues with irregular shapes and large size distributions as well as heterogeneous levels of GUV-GUV adhesion. We show in this work that by achieving GUV assembly in microfluidic confinement we can successfully synthesize prototissues with reproducible shapes and sizes and overall gain control on the GUV-GUV assembly process.  

Microfluidics has emerged as a powerful tool for biomedical research, from cellular to multi-cellular systems \cite{Velve2010,Sackmann2014}.
In particular, the development of minimalistic \textit{in-vitro} organ models  (or \textit{organ-on-a-chip}) and multicellular tumor spheroids in microfluidic platforms has enabled to test their response under controlled chemical and mechanical stresses \cite{Ahadian2017,Yu2010,Aung2023,Wang2023}.
A significant advantage offered by microfluidic techniques is the simplicity in changing the flow geometry by design, together with the simultaneous flow visualization performed with microscopy which permits to relate the material  rheological properties with changes in its microstructure. 
Recent experiments making use of microfluidic micropipette aspiration techniques have been performed in order to probe the rheological properties of cellular spheroids \cite{Tlili2022,Boot2023,Landiech2023} and of individual cells \cite{Davidson2019}. 
Several microfluidic pipette-like devices have been conceived for mechano-phenotyping purposes with the goal of probing simultaneously an ensemble of spheroids \citep{Landiech2023} or of providing a local description of the spheroid rheological features based on the deformation of individual cells and cell-cell rearrangements \cite{Tlili2022}. 
Microfluidic devices can also be advantageous for handling and analysis of synthesized GUVs, by trapping them individually in small wells \cite{Yandrapalli2020} or collectively in larger traps \cite{Yandrapalli2019} or by probing them in micropipette-like devices \cite{Elias2020}. 

We recently reported the first set of microfluidic aspiration experiments on vesicle prototissues \citep{Layachi2022}. 
At the scale of the overall prototissue a viscoelastic behavior was described, which was well captured by a modified Kelvin-Voigt  model (with three independent parameters: elastic modulus, viscosity and relaxation time). 
The rheological description was complemented with velocimetry analysis at the scale of several vesicles. 
We could identify the occurrence of reorganization events taking place between several vesicles by a local increase of the vorticity  field. However, this methodology did not enable to track the deformation nor the displacement of individual vesicles under flow. 
We report in this work the development of a microfluidic platform that makes possible a multi-scale description of model tissues  comprising an analysis at the global scale of the prototissue down to the scale of individual GUVs. 
This becomes possible due to the fact that the assembly of GUV prototissues is achieved in the microfluidic device, which provides prototissues with morphological and imaging properties suitable for image analysis based on segmentation methods.  
In addition, the microfluidic platform enables to drive GUV-GUV adhesion making use of different assembly strategies (based on the streptavidin-biotin binding or the hybridization of complementary DNA strands) and thus to synthesize prototissues with significantly different adhesion strengths. %Prototissues with different levels of adhesion display remarkably distinct flow features both at global and local  scales.  
In addition, the microfluidic platform enables to drive GUV-GUV adhesion making use of different assembly strategies (based on the streptavidin-biotin binding or the hybridization of complementary DNA strands) and thus to synthesize prototissues with significantly different adhesion strengths. %Prototissues with different levels of adhesion display remarkably distinct flow features both at global and local  scales.  

\section{Materials and Methods}
\label{sec:MatMet}
%
%Figure 1:
\begin{figure*}[h!]
\centering
  \includegraphics[height=9.75cm]{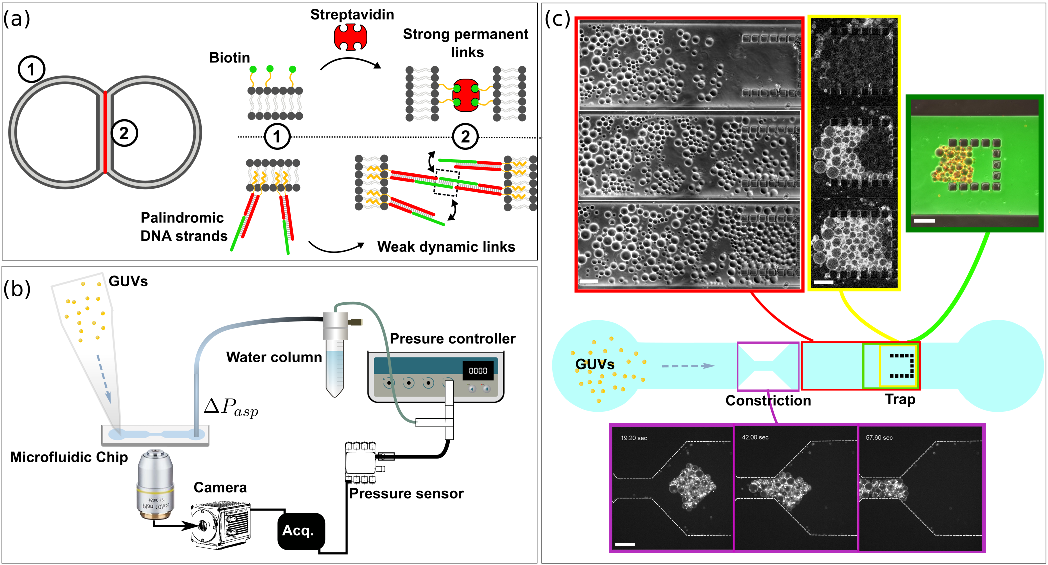}
  \caption{(a) GUV-GUV bonding strategies: an example of a GUV-doublet is shown. Index 1 and 2 refer to the GUV membrane and the GUV-GUV patch respectively. GUV-GUV adhesion can be mediated by the streptavidin-biotin binding (top arrow) or the hybridization of complementary DNA strands (bottom arrow). The DNA anchor and linker sequences are displayed in red and green, respectively. (b) Scheme (not to scale) of the microfluidic device including the microfluidic chip, a device for the control of the applied pressure, a microscope and an acquisition system. (c) Detail of the microfluidic chip used (not to scale) for the prototissue assembly and aspiration tests. Different colors correspond to different parts of the chip used to achieve different tasks: GUV trapping (red), adhesion of GUV-GUV adhesion by the flow of a solution containing fluorescent streptavidin molecules (yellow), release of the prototissue from the trap by reversing the direction of the flow (green) and aspiration test in the microfluidic constriction (purple). The scale bars corresponds to 100 $\mu$m.}
\label{fig:Methods} 
\end{figure*}

\subsection{Prototissue assembly}%: DNA complementary strands \textit{vs.} Streptavidin-Biotin pair}
\label{sec:assembly}
The prototissues analyzed in this work were obtained by the controlled assembly of GUVs.   
GUVs were synthesized by electroformation in a dedicated chamber consisting of two ITO glass slides (SigmaAldrich, 636916) facing each other with a 1 mm thick PDMS spacer. The slides were coated with a lipid solution and left under vacuum to let the solvent evaporate. The lipid mixture was a 9:1 chloroform/methanol solution containing Egg-PC (SigmaAldrich, P3556), DSPE-PEG(2000)-Biotin (Avanti Polar Lipids, 880129P) at 0 $\%$ or 4 $\%$ molar fractions, and ATTO488-DOPE (ATTO-TEC, green marker, $\lambda_\text{ex}=460$ nm, $\lambda_\text{em}=535$ nm) at 1 $\%$ molar fraction. 
GUVs contained a sucrose solution with an osmolarity of 308 mOsm and were dispersed in a glucose solution (with a slightly higher osmolarity $\sim$ 318 mOsm) leading to slightly deflated GUVs. GUVs were stored in a plastic tube at 4 $^{\circ}$C for a maximum of one week. 
In this study, two different approaches were used to drive the assembly of GUVs into prototissues: the incorporation of streptavidin-biotin pairs or complementary DNA strands on GUV membranes. 
This allowed us to obtain two significantly different ranges of GUV-GUV adhesion strengths.

\subsubsection{Streptavidin-biotin mediated GUV assembly~~}
The first method was based on streptavidin-biotin binding, which we already used in our previous studies \citep{Casas-Ferrer2021,Layachi2022}.
Streptavidin (SA) is a tetrameric protein (depicted in red in Fig.~\ref{fig:Methods}-a top row) which can bind specifically to  biotin molecules (depicted in green in Fig.~\ref{fig:Methods}-a top row) which are present in the outer leaflet of the lipid bilayer. 
This binding promotes GUV-GUV bonding through strong permanent links, with an adhesion strength of about 35 $k_BT$ \citep{Chiruvolu1994,Chao2008,Fenz2011,Bihr2014}.
In this study we used a 4 \% molar fraction of biotinylated lipids which corresponded to a surface density on GUV membranes of approximately $\Gamma = 6.15 \times 10^4$ biotin molecules/$\mu$m$^2$. 
Streptavidin (SA) molecules used were marked with a Texas Red fluorophore (ThermoFischer Sci.,S872). The monitoring of SA concentration inside the trap in time was done with fluorescence microscopy. We used 1 $\mu$L of SA solution at 10 $\mu$M for each experiment. 
The amount of SA was significantly larger than the total number of biotin molecules available on the GUVs surfaces contained within the prototissue. 
%Thus the $X$ ratio, which reports the ratio of the total number of molecules of streptavidin in solution to the number of biotin molecules present on the surface of vesicle membranes ($X = n_\text{SA}/n_\text{biot}$) was $X >> 1$ in all experiments. 
Even though SA was in excess in all experiments, saturation of biotin molecules on GUV membranes by SA was prevented (as explained later in Sec.\ \ref{sec:aspiration}). This was possible thanks to the confinement of GUVs into a microfluidic trap, prior to the addition of SA molecules. 
In this configuration GUV interfaces were brought into contact which promoted inter-GUV binding over surface saturation. 
Trap-assisted assembly allowed us to overcome the lack of assembly observed for suspensions of GUVs at high streptavidin-to-biotin ratios due to saturation of biotin molecules by SA \cite{Casas-Ferrer2021}. 

\subsubsection{DNA mediated GUV assembly~~}
The DNA constructs used to drive GUV-GUV assembly were constituted of two main parts: the \textit{anchor} and the \textit{linker}. 
The anchoring part (shown in red in Fig.~\ref{fig:Methods}-a bottom row) was grafted to the outer surface of GUV membranes.  
It consisted of a double stranded DNA (dsDNA). On one end it was functionnalized with two cholesterol molecules to allow insertion into the lipid bilayer (represented in orange in the sketch) \cite{Pfeiffer2004}. On the other end, the construct had a single stranded (ssDNA) ending which enabled linker attachment. The anchor was marked with a red fluorophore (Atto-565). 
The linker consisted of a single stranded (ssDNA) oligomer, complementary to the anchor strand on its $3'$ end, and with a short palindromic sequence on the $5'$ end. Palindromic sequences are self complementary and can hybridize to themselves. After hybridization of the linker, GUVs became adhesive due to the hybridization of the palindromic domain at their surface. The linker was marked with a green Atto488 fluorophore. A sketch showing the DNA binding mechanism is shown in Fig.~\ref{fig:Methods}-a. 
The strength of the adhesion was set by the number of complementary bases of the palindromic domain, which is shown with a squared region in the sketch of Fig.~\ref{fig:Methods}-a (bottom row). 
In this study we used a palindromic domain with eight bases, which provided a weak adhesion strength of about $\sim 3 k_B T$ per binding module (further details about the thermodynamic data of the DNA strands are provided in the ESI\dag).  
Despite the low adhesion between DNA strands, at the GUV-GUV interface the collective hybridization of multiple DNA duplexes was sufficient to induce adhesion between GUVs and lead to the successful formation of cohesive GUV prototissues.  

All DNA strands were reconstituted from a lyophilized commercial preparation (from Eurogentech) in TE buffer (containing 10 mM Tris and 1 mM EDTA, from Sigma Aldrich, and adjusted to pH=8).  
The anchor was assembled in the thermocycler by annealing the mixture to $85 ^{\circ}$C and cooling it down to $25 ^{\circ}$C, at $0.1 ^{\circ}$C/s. 
This was achieved using two equimolar solutions (10 $\mu$M) of the complementary strands in a TE buffer containing 150 mM of NaCl. 
GUVs were first functionalized with the DNA-anchor in order to saturate the outer leaflet of the membrane of the GUVs. 
For this, we mixed 100 $\mu$L of the GUVs suspension at a concentration of approximately $4 \times 10^3$  GUV/$\mu$L with 5 $\mu$L of the anchor solution at 10 $\mu$M and 100 $\mu$L of TE-NaCl buffer solution in a low binding tube. The GUV solution was then left to incubate for 2 hours at room temperature on a rotatory mixer to allow homogeneous grafting of the membranes. The excess anchor present in the GUV solution was removed by performing several centrifugation steps. We estimated the GUV surface coverage to be around 1000 strands/$\mu$m$^2$, based on a calibration step which allowed us to relate fluorescence intensity values with DNA concentrations (data not shown). 
The DNA-linker was then incorporated into the suspension of functionalized GUVs in order to mediate the adhesion between GUVs. The concentration of linker was sufficient to hybridize all anchoring strands present on GUV membranes. 
This step was achieved in the microfluidic trap. As mentioned earlier, since GUVs were highly confined prior to the addition of the linker molecules, this assembly protocol maximized GUV-GUV adhesion patch formation (and saturation of DNA strands on GUV membranes was prevented).

\subsection{Microfluidic platform for prototissue assembly and testing}
\subsubsection{Microfluidic apparatus~~}
The microfluidic geometry consisted of a 200 $\mu$m wide channel leading to a $100\pm5$ $\mu$m wide constriction (a sketch is shown in Fig.~\ref{fig:Methods}-c). 
The decrease in width followed an angle of 45 $^{\circ}$ to avoid strong recirculation zones at the edges. The constriction was 200 $\mu$m long and was considered small compared to the total channel length (4 mm). 
All channels had a height of $50\pm7$ $\mu$m. 
On one side of the channel, a built-in trap consisting of a series of micrometric squared pillars (40 $\mu$m wide with a spacing of 15 $\mu$m) was designed to retain GUVs, while letting water flow, and accumulate them. This allowed us to synthesize prototissue with controlled shapes and sizes. 
Microchannels were fabricated by standard soft-lithography microfabrication techniques \citep{Tabeling2005}. A negative photoresist (AZ125nXT, Merck) was spincoated onto a silicon wafer. The pattern was lithographed onto the photoresist using a maskless lithography beamer (SmartPrint). After curing and development, the devices were molded using polydimethylsiloxane (PDMS) and sealed to a glass microscope slide using plasma bonding. Prior to all experiments, the channels were functionalized with a $\beta$-casein solution to reduce friction of GUVs in contact with the channel walls \citep{Lee2022}. 
The flow inside the chip was generated by a pressure difference (Fig.~\ref{fig:Methods}-b). A height-adjustable water column was connected to the outlet of the microfluidic chip to provide a positive pressure difference. In addition, the water column was connected to a pressure controller (Fluigent, MFCS-EZ) providing a negative pressure (with a resolution of 2 Pa). With the tracking of fluorescent micro-beads (Invitrogen, Molecular Probes), both the height of the water column and the pressure controller offset were adjusted to immobilize the beads inside the channel and thus to set the pressure to zero before starting any microfluidic experiment. 

\subsubsection{Assembly and aspiration protocol~~}
\label{sec:aspiration}
The microfluidic chip was designed to be able to achieve three main tasks: (i) accumulate the GUVs in the microfluidic trap with tailored size and shape, (ii) mediate the assembly of GUVs into a prototissue (making use of streptavidin-biotin binding or DNA strands), and (iii) perform the rheological assay of the prototissue by microfluidic aspiration.
For this, we first plugged a reservoir filled with a suspension of GUVs in one side of the chip (left inlet in Fig.\ref{fig:Methods}-c). 
We let the GUVs sediment at the channel entrance, and we then applied a gentle negative pressure difference in order to flow the GUVs towards the trap and accumulate them in it (red frame in Fig.\ref{fig:Methods}-c). 
As soon as a sufficient number of GUVs was accumulated in the trap, we injected the binding solution (containing fluorescent streptavidin or DNA-linkers) through the inlet. GUV-GUV bonding was achieved once the binding solution reached the GUVs contained within the trap. An example is shown in the time lapse displayed in the yellow frame in Fig.\ref{fig:Methods}-c in which the advection of streptavidin (marked with a fluorescent marker) inside the trap can be monitored.  
Once all GUVs were fonctionalized with the linker molecule we reversed the direction of the flow in order to release the formed prototissue from the trap (green frame in Fig.\ref{fig:Methods}-c). 
We then applied a gentle flow to bring the prototissue close to the channel constriction in order to perform its rheological characterization (purple frame in Fig.\ref{fig:Methods}-c).
The protocol we used for the microfluidic rheology experiments consisted of steps of increasing pressure values which enabled to deform the prototissue progressively, in a controlled fashion. In most of the described experiments we applied steps of 10 Pa during 10 s from 0 to 70 Pa. 
We controlled the applied pressure making use of a Fluigent pressure controller and the Oxygen software. 
The applied pressure was measured independently using a pressure sensor (Honeywell) with a resolution of 0.5 Pa in order to double check the value provided by the pressure controller. 
As an example, two videos corresponding to a microfluidic aspiration experiment obtained using a SA-biotin and a DNA- mediated prototissue are included in the ESI\dag.   

\subsection{Image acquisition and analysis}
\label{sec:imaging}
We achieved the visualization and monitoring of the microfluidic experiments making use of phase contrast an fluorescence microscopy. 
We used an inverted microscope (Leica, DMi8) with a $20\times$ magnification air objective ($\text{NA}=0.4$).
Fluorescence excitation was provided using a white LED lamp (CoolLED, pE-300) coupled to fluorescence filters during a 30 
ms exposure time. Images were captured at 30 fps with a CMOS camera (Hamamatsu, C13440 ORCA flash 4.0). 
Fluorescence images were analyzed using ImageJ homemade routines in order to identify and quantify the properties of individual GUVs, pairs of GUVs or GUV-prototissues. 
\subsubsection{Deformation of individual GUVs~~}
We developed an ImageJ routine with the aim of characterizing the shape of individual GUVs. 
Fluorescence images of GUVs were analyzed using intensity based image segmentation methods in order to identify the contours of individual GUVs. We quantified the shape of GUVs using two different shape descriptors. 
First, we defined an \textit{angularity} ($\Phi$) descriptor which allowed us to distinguish circular (smooth) GUV contours from polygonal (angular) ones.  
This descriptor quantifies the departure of a vesicle contour from a circular shape, by computing the number of bright pixels on the contour lying away from an ideal circle. This descriptor was useful to successfully monitor the binding process between adjacent GUVs contained within the microfluidic trap, since GUVs transitioned from circular (with $\Phi\simeq 0$) to polygonal shapes (with $\Phi> 0$) as they became adhesive. More details on the calculation of the angularity as well as examples obtained for GUVs with different shapes can be found in the ESI\dag.   
Second, we fitted the contour of individual GUVs with an ellipse and used the value of the ellipse eccentricity ($\epsilon$) as a proxy for the deformation of individual GUVs. 
The eccentricity was computed as the ratio of the distance between the foci of the ellipse to its major axis length. The eccentricity can range from 0 to 1 depending on the aspect ratio of the ellipse (0 and 1 are degenerate cases corresponding to a circle or a line segment, respectively). 
\subsubsection{Tracking of GUVs and inter-GUV displacements~~}
We tracked individual GUVs in time using the FastTrack software \cite{Gallois2021}. 
This allowed us to reconstruct the trajectories of individual GUVs contained within the prototissue all along an aspiration experiment and map the value of their deformation ($\epsilon$) in time.   
Monitoring of GUVs over time also allowed us to quantify spatial reorganizations of GUVs taking place within the prototissue. 
For this, we defined a mesh based on the initial positions of a set of individual GUVs contained in the prototissue. 
The mesh was constructed by applying Delaunay triangulation to the $x$ and $y$ coordinates of the center of each GUV.  
We then defined for each pair of GUVs ($i,j$) the initial distance between them ($d_{ij,0}$). 
The relative distance between GUVs constituting each pair ($d{ij,t}$) was computed in time. 
In order not to take into account the initial value of the distances, which could be very different among the different pairs of vesicles, the distance was normalized by the initial inter-GUV distance as
as $\gamma_{ij,t}=d_{ij}/d_{ij,0}$. We name this variable hereafter \textit{inter-GUV displacement}. 
The inter-GUV displacement can be interpreted as a proxy of the local strain along the GUV-GUV axis displayed by the prototissue.   
Since the prototissues were not deformed isotropically in the channel constriction, this method allowed us to map the strain locally. 
An example of the mesh is shown in Fig.\ \ref{fig:Lagrange}-c, and it will be discussed further in Sec.\ \ref{sec:local}.
Complementary to it, we measured the deformation of individual GUVs. As described earlier we fitted an ellipse to each GUV and computed its eccentricity ($\epsilon$). 
We defined an \textit{intra-GUV} deformation for each pair of GUVs based on the mean eccentricity of the two GUVs as $\epsilon_{ij}=(\epsilon_{i}+\epsilon_{j})/2$. 
Again, we computed the relative variation with respect to the initial value and made it dimensionless as $\Sigma_{ij,t}=\epsilon_{ij}/\epsilon_{ij,0}$. 
The intra-GUV can be interpreted as a proxy for the local stress felt by individual GUVs.  

\subsubsection{Deformation of vesicle prototissues~~} 
Finally, we aimed at monitoring the advancement position as well as the shape of the entire prototissue during the aspiration experiments. 
For this we applied the same image treatment developed in our previous work \citep{Layachi2022}. 
Briefly, we used an ImageJ routine to detected the contour of the prototissue (that we consider now as a single object) based on intensity-based image segmentation. Segmented images were analyzed using a Matlab homemade routine to extract the contact length ($l_c$) defined as the front position of the prototissue inserted inside the channel constriction and in contact with the lateral  walls. 
The strain (dimensionless deformation) was computed as the variation of the contact length relative to its initial value, as $\gamma_{Tissue}= (l_c-l_{c,0})/l_{c,0}$. 
The prototissue deformation was monitored in time during the aspiration steps and used to compute the values of both the viscosity,  elastic modulus and relaxation time of the prototissue.  

\section{Results and Discussion}
\subsection{Microfluidic conception of prototissues with tailored mechanical properties}
The microfluidic chip was designed to enable the assembly of GUV prototissues \textit{in situ}. Such design made possible the conception of prototissues with tailored morphological properties (size and shape) and adhesion properties.  
The analysis of the \textit{static} properties of the prototissues contained inside the microfluidic trap is addressed in the following section.   

\subsubsection{Static properties~~}
\begin{figure}%[h!]
\centering
  \includegraphics[height=13.5cm]{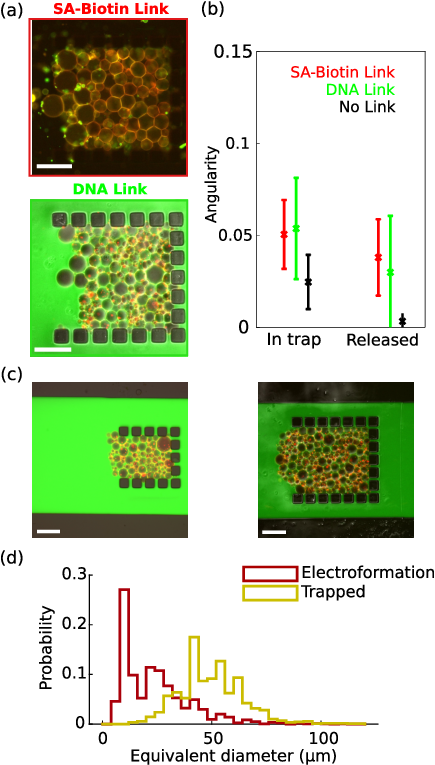}
  \caption{(a) Example of two GUV-prototissues obtained in the microfluidic trap using SA-biotin binding (red) or DNA complementary strands (green). (b) Angularity ($\Phi$) values obtained for an ensemble of GUVs. Different colors correspond to different adhesion molecules (SA-biotin in red, DNA in green and black in the absence of linker molecules). The values corresponding to GUVs contained within prototissues located inside the trap or after being released are compared. The displayed values correspond to the mean obtained over 10 GUVs (located away from the trap borders) and the error bars to the standard deviation. (c). Example of different prototissues obtained using microfluidic traps of different sizes (left 180 $\mu$m $\times$ 220 $\mu$m and right 290 $\mu$m $\times$ 330 $\mu$m). (d) Size distribution of the GUVs obtained after electroformation (red) and of the GUVs contained in the microfluidic trap (yellow). The scale bars correspond to 100 $\mu$m.}
  \label{fig:Methods2}
\end{figure}

We analyzed the degree of cohesion of prototissues in the microfluidic trap by performing a morphological analysis of the vesicle contours. 
For this we used the angularity shape descriptor ($\Phi$). As discussed in Sec. \ref{sec:imaging}
, the angularity is sensitive to the existence of sharp angles along the object contour.
Since GUV contours transitioned from non-adhesive (with mainly circular shapes) towards adhesive GUVs (with polygonal shapes) we could use $\Phi$ to monitor the formation of GUV-GUV adhesion patches between adjacent GUVs contained inside the microfluidic trap. 
Likewise, the angularity could be used as a suitable shape descriptor to quantify the degree of the prototissue cohesion. 
In Fig.\ \ref{fig:Methods2}-a we include an example of two vesicle prototissues contained inside the microfluidic trap obtained with SA-biotin or DNA complementary strands. 
In panel (b) we provide the values of the angularity computed for an ensemble of GUVs (belonging to different prototissues with comparable adhesion strengths).  
The values of $\Phi$ obtained for the SA-biotin and DNA mediated assemblies are shown in red and green respectively. 
There is no significant difference between the two conditions.  
In both cases GUVs were slightly deflated and thus their membranes were easily deformable upon binding \cite{Ramachandran2011} and their contour displayed  polygonal shapes. 
Besides, it is visible in Fig.\ \ref{fig:Methods2}-b that in the absence of linker molecules (shown in black) the angularity of GUVs was not zero in the trap. This was due to the presence of an outer pressure field that we applied in order to keep the GUVs inside the trap. Although the resulting flow was gentle, it was sufficient to deform the GUV membranes. 
In panel (b) we also compare the angularity of GUVs when they were kept inside the microfluidic trap and after they were released. 
We see that in the absence of linker, once the GUVs were released from the trap by reversing the flow direction, angularity values dropped down to zero since GUVs recovered an almost circular shape after they were released. 
However, when GUVs were adhesive the values of angularity after release only diminished slightly and remained positive. 
This is indicative of a persistent GUV-GUV adhesion mediated by SA-biotin or DNA strands.  
The variation in angularity, displayed by the error bars, comes from differences in the local environment of each GUV (given by a different number of neighboring GUVs and the location of each GUV within the prototissue close or far away from the edge).   
Morphological descriptors (such as the angularity) may not be however sufficient to discriminate between different levels of adhesion. 
Once an ensemble of GUVs reaches a maximal packing, an increase in the adhesion strength may not longer result into a morphological change. In order to characterize the differences in adhesion strength we made use of rheological descriptors, which inform about the response of the systems under flow. As it will be described in Sec.\ \ref{sec:multiscale}, rheological descriptors are sensitive to  different adhesion strengths, which were provided by the use of SA-biotin binding or DNA strands.  

The microfluidic platform we designed also allowed us to tailor the size and shape of the assembled prototissues, which was set by the size and shape of the microfluidic trap itself.  
We show an example of two prototissues with nearly squared shapes obtained with traps of different lateral sizes (180 and 290 $\mu$m).  
After release from the trap, the shape of the prototissues was mainly preserved. This is visible in Fig.\ \ref{fig:Methods}-c in which we can appreciate that even the sharp corners were preserved after release.    

An additional advantage of designing the vesicle prototissue inside the microfluidic device, compared to a bulk assembly protocol, was the decrease in polydispersity of the GUVs contained inside the prototissue. 
GUVs were fabricated making use of electroformation. This is a very efficient technique as it provides large amounts of GUVs ($\sim 10^6$ GUVs per electroformation) but it provides GUVs which display an important polydispersity. 
The microfluidic protocol we used enabled to reduce significantly this polydispersity, as it permitted to filter out the smallest GUVs. 
A histogram comparing the size distribution of GUVs obtained by electroformation and the GUVs immobilized inside the trap is shown in Fig.\ \ref{fig:Methods2}-d. 
This filtering process was effective, in part, at the GUV reservoir placed at the microchannel entry. As discussed in Sec.\ \ref{sec:aspiration}
 we let GUVs sediment in order to reach the channel. Since small GUVs took a longer time to sediment they did not get  inserted inside the channel. 
The filtering process was enhanced at the microfluidic trap. The trap was constituted of pillars with a small spacing of  15 $\mu$m  between them which enabled the filtering of the smallest GUVs contained in the trap.

\subsection{Multiscale rheological description}
\label{sec:multiscale}

%Figure 3: Euler
\begin{figure}%[h!]
\centering
	  \includegraphics[height=18cm]{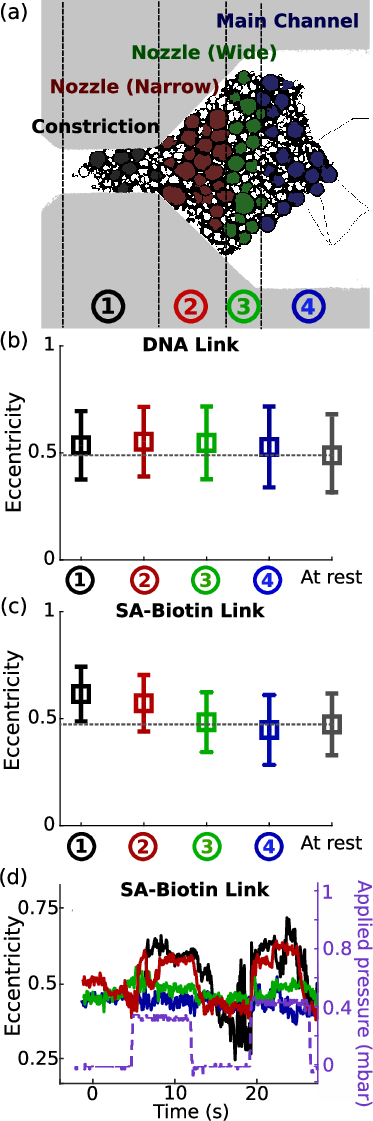}
  \caption{(a). Snapshot of the microfluidic channel. The identification of individual vesicles contained within the prototissue was achieved based on image segmentation methods. Vesicles belonging to different regions of the channel are labeled with different colors: (1) constriction (black); (2) narrower part of the nozzle, next to the constriction (red); (3) wider part of the nozzle, next do the main channel (green); (4) main channel (blue). (b-c). Mean eccentricity values obtained for each channel region after averaging  over all the GUVs located in a given region and over time. The error bars corresepond to the standard deviation. GUV-GUV  assembly was mediated by DNA (panel (b)) or SA-biotin binding (panel (c)). (d). Evolution of the mean eccentricity of GUVs over time, obtained during an aspiration sequence for a SA-biotin-prototissue. Each color codes for a different region. The applied pressure signal is also displayed with a purple dashed line (right axis).}
\label{fig:Euler}
\end{figure}
We provide in this section a comprehensive rheological description of prototissues including a local description, at the scale of individual GUVs, and at the global scale of the entire prototissues.   
In microfluidic aspiration experiments, prototissues were mechanically deformed as they got inserted into the channel constriction by the application of an external pressure difference. 
At the local scale, the application of the pressure difference also resulted into the deformation of individual GUVs contained within the prototissue and to GUV-GUV spatial reorganizations.
The multi-scale approach that we perform in this section enables to understand the relationship existing between these two different length scales. 

\subsubsection{Local scale: individual GUVs~~}
\label{sec:local}
In order to quantify the evolution of the deformation of GUVs under flow we used an Eulerian approach with a fixed stationary reference frame. In order to facilitate the analysis, we divided the channel in four distinct regions (constriction, narrow nozzle, wide nozzle and main channel), which are highlighted with  different numbers and colors in Fig.\ \ref{fig:Euler}-a. 
We used the eccentricity in order to quantify the deformation of individual GUVs contained within the prototissue, as detailed in Sec. \ref{sec:imaging}.  
The eccentricity of all vesicles contained in each channel region was measured all throughout the aspiration phase (during which a positive pressure difference was applied) and was averaged in time ($t$) and for all vesicles passing through each region. The data shown in Fig.\ \ref{fig:Euler}-b correspond to the average eccentricity values obtained in the different channel regions, $\left\langle \epsilon\right\rangle_{Region,t}$. Vesicle deformation was also measured at rest for comparison (shown in grey in Figs.\ \ref{fig:Euler}-b,c). 
Note that the typical values obtained for vesicles at rest, which have nearly round shapes, are $\epsilon \simeq 0.5$. 
Although $\epsilon \simeq 0$ would be expected for spherical objects, these relatively large values may result as a consequence of the segmentation process which introduces slight morphological perturbations of the object contours, as compared to ideal round shapes. We thus compare the eccentricity values obtained for GUVs to this reference value (that we have highlighted in the figure with a dashed line).  
It is visible from the figure that for DNA-mediated prototissues there is no significant difference between the eccentricity of the GUVs located in the different channel regions. However, in all regions the mean eccentricity is slightly larger than the value of the mean eccentricity obtained for the prototissue at rest, in the absence of flow. 
This implies that GUVs were only very slightly deformed under flow for the case of weakly-adhesive DNA-mediated prototissues, even in the channel constriction. 
On the contrary, for SA-biotin-mediated prototissues the mean eccentricity values increase gradually for GUVs located closer and closer to the channel constriction. 
Thus, for these highly-adhesive prototissues GUVs got deformed along the flow streamlines as they approached the constriction channel. 
In Fig.\ \ref{fig:Euler}-d the temporal evolution of the mean eccentricity in the different channel regions is assessed for a SA-biotin-mediated prototissue. 
It is visible that the mean deformation in each region is in-phase with the applied pressure (shown with a dashed purple line in the figure). 
An increase of the mean eccentricity takes place during each step in pressure. 
These significant changes in the eccentricity values are observed mainly in regions 1 and 2, corresponding to the constriction and the narrow-nozzle region next to it. 
In region 3 (wider nozzle region) the deformation is less important and it becomes almost negligible in region 4 (main channel) far away from the constriction. 

%Figure 4: Lagrange
\begin{figure*}[h!]
\centering
  \includegraphics[height=8cm]{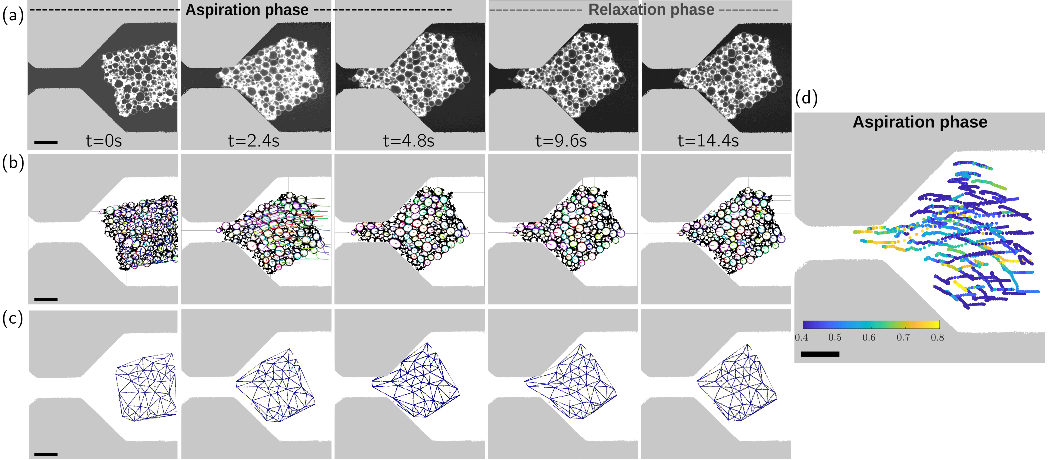} 
	 \caption{(a) Representative snapshots of a microfluidic aspiration experiment of a vesicle prototissue obtained with SA-biotin binding, including the aspiration and relaxation phases. (b) Identification and tracking of individual vesicles contained within the prototissue obtained using the FastTrack software and a home-made image segmentation routine. Each GUV is represented with a different color. (c)  Geometrical mesh used to determine the inter-GUV displacement. Vertexs of the mesh correspond to the center of each GUV contained in the prototissue, and dots highlight the mid-position of all GUV-GUV distances (represented in blue). (d) Spatiotemporal map displaying the eccentricity values obtained for GUVs constituting the prototissue over an aspiration phase. The scale bars in all panels correspond to 100 $\mu$m.}
  \label{fig:Lagrange}
\end{figure*} 

%\paragraph*{Fixed \textit{vs.} co-moving frameworks.~~} 
\paragraph*{Eulerian \textit{vs.} Lagrangian frameworks.~~} 
We complemented the analysis at the local scale by tracking individual vesicles in time. This analysis corresponds to a Lagrangian approach for which the reference frame is not fixed but moves with the object under study. 
In Fig.\ \ref{fig:Lagrange} we show an example of a microfluidic aspiration experiment obtained with a SA-biotin-mediated prototissue.  
In panel (a) five selected frames obtained by fluorescence microscopy are shown at different times, covering the aspiration phase during which the pressure difference was applied (first 3 frames) and the relaxation phase during which the pressure was released (2 last frames).  
In panel (b) the identification of individual GUVs contained within the prototissue resulting from the segmentation analysis is illustrated for the same five frames. 
Individual GUVs are highlighted with different colors and the individual trajectory of each GUV center is tracked in time. 
In the figure, and for the sake of clarity, we only represent the position of each GUV over ten frames. 
The ellipses fitted to each GUV contours and used to computed GUVs eccentricity are also shown. 
In panel (d) GUV eccentricity values are displayed in the form of a spatio-temporal map, corresponding solely to the aspiration phase.  
It is visible from the map that the values of eccentricity increase when GUVs approach the constriction region. 
This is in agreement with the description provided using the Eulerian approach. 
Note, however, that GUVs placed at the rear edge of the prototissue display in general small eccentricity values (corresponding to very slightly deformed GUVs) throughout the entire aspiration phase. This could be attributed to the fact that these GUVs are surrounded by a fewer number of neighboring GUVs, compared to GUVs found in the center of the prototissue. This partial confinement would translate into a reduced GUV deformation under flow. 
These analysis shows the complementarity of both the Eulerian and Lagrangian approaches. 
While the large eccentricities measured at the vicinity of the constriction are visible with both approaches, only the Lagrangian one enables to describe the role of GUVs microenvironments in the eccentricity.

\paragraph*{Local micro-rheology measurements.~~} 
The local response of vesicle prototissues after a step in pressure was viscoelastic, as it combined an elastic and a viscous contribution. 
As we have already discussed, part of the strain results from the elastic deformation of individual GUVs contained within the prototissue.  
The other part of the strain results from dissipative reorganization events taking place between GUVs contained in the prototissue. Other dissipative processes include viscous dissipation of the outer Newtonian aqueous solvent, as well as friction of the prototissue with the channel walls. 
However, since the viscosity of the solvent ($\eta_s = 1.3$ mPa$\cdot$s) is negligible compared to the viscosity of GUV prototissues (as it will be described later), the solvent contribution can be disregarded. 
Likewise, in the microfluidic experiments we took care of reducing the friction term with the channel walls by passivating the microfluidic channel walls with a casein solution prior to any experiment. 
Thus we can consider the spatio-temporal reorganizations of GUVs within the prototissue as the main source of viscous dissipation, which we analyse in the following. 

In order to quantify GUV-GUV reorganizations taking place within a prototissue under flow we defined a 
geometrical mesh by linking the center of neighboring GUVs, which allowed us to set and monitor the distance between GUVs. The 2D-mesh was computed based on the position of more than 50 GUVs.  
The mesh is represented in Fig.\ \ref{fig:Lagrange}-c for five selected frames (GUVs are no longer shown for clarity reasons). 
Each vertex of the mesh corresponds to the center of each GUV, and each blue segment to the link between two GUVs. It is visible in the figure that the length of the mesh segments increases significantly in time as the prototissue is driven inside the constriction (and diminishes in the relaxation phase). These variations are more significant for the GUV-pairs located close to the constriction and they are almost negligible for the GUV-pairs far apart from it.  
We compute the displacement between pairs of GUVs (that we name \textit{inter-GUV displacement}, $\gamma_{ij}$) as the relative variation of the GUV-GUV distances in time (as previously described in Sec.\ \ref{sec:imaging}). 
%
%Figure 5: local-pairs
\begin{figure}[t!]
\centering
  \includegraphics[height=10.5cm]{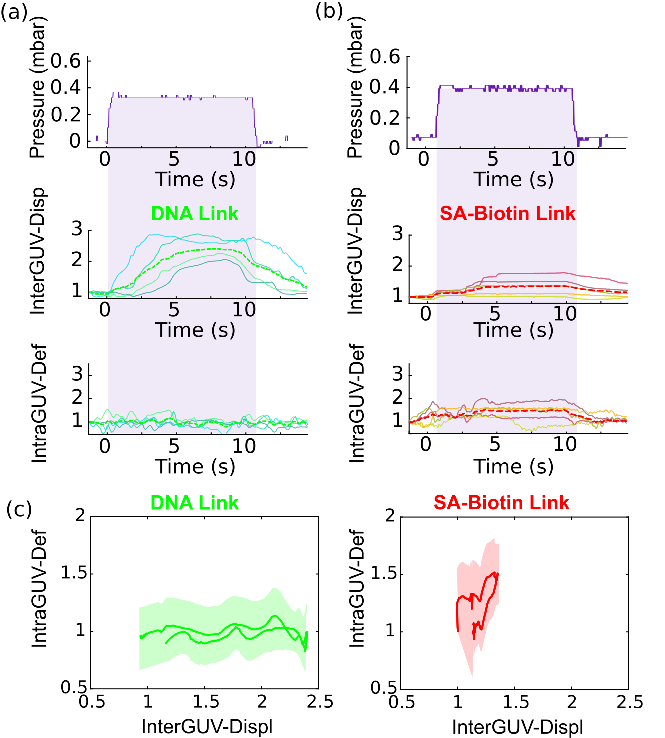}
  \caption{(a-b) 	Pressure (top panels), inter-GUV displacements (medium panels) and intra-GUV deformations (bottom panels) as a function of time. Different curves (shown in different colors) correspond to five different pairs of GUVs. Mean values are shown with a dashed line. Panel (a) corresponds to a prototissue assembled with DNA strands and (b) with SA-biotin binding. (c) Intra-GUV deformation represented as a function of the inter-GUV displacement for DNA and SA-biotin prototissues. The area explored by five different GUV-pairs is shown with shaded regions and the mean is represented with a solid line.}
  \label{fig:pairs}
\end{figure}
In Fig.\ \ref{fig:pairs} we represent the evolution of the inter-GUV displacement in time, corresponding to four different pairs of GUVs located close to the channel constriction. The mean of the four pairs is shown with a dashed line.  
In order to complete the rheological characterization of the prototissue at the local scale we also show the temporal evolution of the intra-GUV deformation ($\Sigma_{ij}$) corresponding to the same pairs of GUVs, computed based on the relative variation of eccentricity (as described in Sec.\ \ref{sec:imaging}). 
Inter-GUV displacements can be interpreted as a source for viscous dissipation (accounting for GUV-GUV rearrangements), while intra-GUV deformations can be interpreted as an elastic response of the material.  
We also include in the figure the pressure signal as a function of time, for which the aspiration phase has been highlighted with a shaded region.
The results obtained for DNA and SA-biotin-mediated prototissues are shown, respectively, in panels (a) and (b). 
These two different types of prototissues display qualitatively different flow behaviors after a step in pressure, both in terms of  inter-GUV and intra-GUV responses.  

The DNA-mediated prototissue shows a flow response dominated by an increase of inter-GUV displacements, which can reach up to three times their initial values. 
The DNA linkers we used have a weak adhesion of few $k_B$T. 
Thus, GUV-GUV bonds mediated by DNA strands could break locally when prototissues were submitted to external extensional or shear stresses. The dynamic nature of DNA bonding allowed GUV-GUV adhesion sites to rebuild once the stress was released. 
The application of an external stress could also lead to the detachment of DNA strands from the GUV lipid bilayer.  
However since DNA anchoring into the lipid bilayer was mediated by two cholesterol moieties, detachment of DNA from the membrane was unlikely \citep{Pfeiffer2004,Bennett2009}. 
The variation of the intra-GUV deformation observed in Fig.\ \ref{fig:pairs}-a for the DNA-mediated prototissue is instead very small, as already discussed  previously. We believe that the adhesion mediated by DNA strands was too low to deform individual GUVs.  
For such low-adhesive GUV-GUV contacts the external geometrical compression mediated by the channel nozzle was not efficiently transmitted through the prototissue leading to negligible deformations of GUVs. 
As shown in the figure, for most GUV-pairs the initial values of inter-GUV displacements are partially recovered after pressure release. 
Note that total recovery would be expected for an ideal laminar Stokes flow for which identical trajectories should be obtained for back and forth forcing, as the application and removal of the pressure corresponds to a symmetric time-reversal protocol.  

The SA-biotin-mediated prototissue, instead, shows a response that results from a combination of both inter-GUV displacements as well as intra-GUV deformation (panel (b)).  
The magnitude of the inter-GUV displacements is smaller compared to those observed for the DNA-mediated prototissue. 
On the contrary, the magnitude of the intra-GUV deformation of SA-biotin-mediated prototissues is significantly larger.
The adhesion patches between GUVs mediated by SA-biotin ensured the cohesion of the prototissue under flow. These made possible the translation of the compression strain, exerted geometrically by the narrowing walls of the constriction channel, into an increase of GUV eccentricity. 
For most GUV-pairs the intra-GUV deformations diminish significantly after the pressure release, which accounts for the elastic recovery towards their initial shapes.
GUV-GUV adhesion mediated by SA-biotin is strong making unfavorable the rupture of the binding SA-biotin pair under stress. Thus the rupture of GUV-GUV bonds should be most likely facilitated by the pull-out of biotinylated-lipids from the lipid bilayer. 
In such scenario, GUV-GUV detachment should be considered as irreversible, since detached biotinylated lipids from vesicle membranes would be more easily reinserted into the lipidic bilayer of the same GUV carrying the SA-biotin binding pair (due to their immediate proximity) rather than the membrane of a neighboring GUV, and would thus not lead to any new GUV-GUV link. 
Likewise, since GUV surfaces were saturated with SA, the eventual rupture of SA-biotin bonds was also irreversible. 
In panel (c) we represent the intra-GUV deformation as a function of the inter-GUV displacement for both types of prototissue. 
The shaded line corresponds to all the region explored by the different pairs of GUVs and the continuous line is the mean. 
It is visible that the trajectory of the DNA-mediated prototissue in this representation expands only along the inter-GUV displacement (horizontal axis). The deformation of GUVs remains negligible regardless of the local strain at which the prototissue is subjected. 
Instead, the results corresponding to the highly adhesive prototissue, mediated by SA-biotin binding, expand equally along both the \textit{inter}- and \textit{intra} -GUV related deformation axis. 
There exists a correlation between both variables: small values of inter-GUV displacements correspond to small values of intra-GUV deformations, and similarly for larger values. 
Under an imposed external forcing, GUV-GUV adhesion patches are strong enough to build up internal stresses within the prototissue and enable the elastic deformation of GUVs. Elastic deformations increase for larger forcings. But the deformation of individual GUVs comes along with GUV-GUV reorganizations, since GUV-GUV patches can eventually rupture at large forcings. 
This is in agreement with previous results obtained for other complex materials under flow, such as granular materials, foams or emulsions. 
Both the rate of exchange and the deformation of the subunits have been shown to depend on the cohesion of the material (given partly by the liquid content in foams \cite{Cohen-Addad2013,Dollet2014}, or the degree of adhesion between droplets in adhesive emulsions \cite{Golovkova2020,Montel2022}). 
The combination of both modes of deformation, reversible and irreversible, has been modeled theoretically with different viscoelastic \cite{Guevorkian2010,Tlili2020,Tlili2022}, elastoplastic \cite{Golovkova2020} or viscoelastoplastic models \cite{Tlili2015,Dollet2010}, according to the specific flowing trends exhibited by each material.    
In our previous work \cite{Layachi2022} we showed that the flow of vesicle prototissues could be well captured by a viscoelastic 
 Kelvin-Voigt model with three independent parameters. This modeling part at the global scale is addressed in the following section. 

\subsubsection{Global scale: GUV prototissue~~}
%Figure 6: Global
\begin{figure*}[h!]
\centering
\includegraphics[height=7cm]{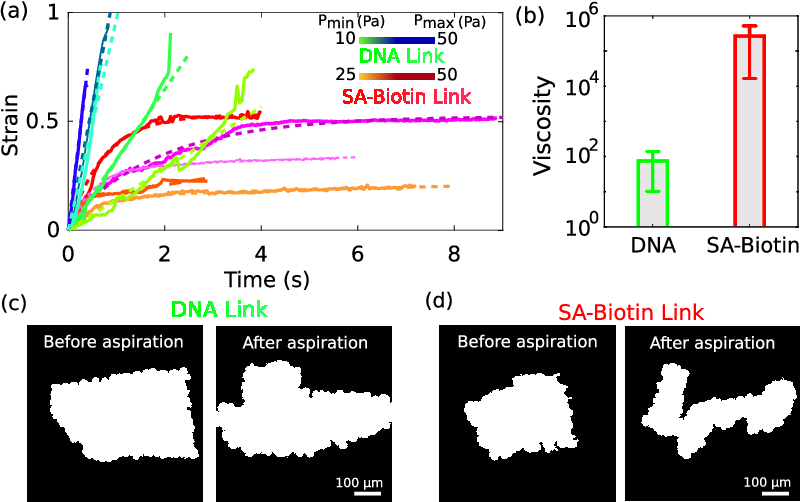}
  \caption{(a). Prototissue strain ($\gamma{Tissue}$) evolution as a function of time obtained after an aspiration experiment for a prototissue mediated by DNA strands (green to blue colors) or SA-biotin binding (orange to red colors). Different curves correspond to different applied pressures (b). Mean viscosity values obtained for the DNA (green) and SA-biotin prototissues (red). The error bars correspond to the standard deviation. (c-d). Contour of a prototissue obtained before and after performing the microfluidic experiment for a DNA (c) and a SA-biotin (d) mediated prototissue.} 
  \label{fig:global}
\end{figure*}
We complement the rheological characterization of vesicle prototissues by performing an analysis at the global scale of the entire prototissue. 
We base this analysis on the measurement of the advancement of the front of the prototissue in time inside the microfluidic constriction. 
In Fig.\ \ref{fig:global}-a we show the strain of a prototissue ($\gamma_{Tissue}$) mediated by DNA strands (green) or SA-biotin binding (red) as a function of time. Different curves correspond to different values of the applied pressure.  
The strain increase measured for the DNA-mediated prototissue is fitted with a linear regression (shown with dashed lines). A  linear trend would correspond to the response of a purely viscous behavior in a creep experiment, for which $\gamma (t) = \dot{\gamma} (t-t_0)$ ($t_0$ corresponds to the initial time for each step in pressure and $\dot{\gamma}$ to the shear rate). 
The linear fit shows good agreement with the experimental data. 
The viscosity ($\eta$) can be determined from the slope of the linear fit and the applied pressure as $\eta \sim \Delta P/\dot{\gamma}$. For a purely viscous behavior the elastic modulus and relaxation time are identically equal to zero. The values of viscosity obtained for the DNA-mediated prototissue range within 5 and 150 Pa$\cdot$s (the mean and standard deviation are displayed in panel (b)). 
Instead, the strain evolution measured for the SA-biotin-mediated prototissue displays an initial fast
exponential increase followed by a slower regime for which the strain increases linearly in time. 
This flow response is typical of viscoelastic materials, which exhibit an elastically-dominated behavior at
short times and a viscous one at longer times. 
The strain curves can be fitted using a modified Kelvin-Voigt model, which contains an initial exponential phase and a subsequent linear increase: $\gamma (t) = \gamma_0\left\{1-\exp\left[-(t-t_0)\right]/\lambda\right\}+\dot{\gamma}_{\infty} (t-t_0)$ (where $\gamma_0$ and $\dot{\gamma}_{\infty}$ are the intercept and the slope at long times, respectively). This viscoelastic model contains three independent parameters: viscosity ($\eta$), elastic modulus ($G$) and relaxation time ($\lambda$).  
The viscosity can be obtained from the fitting parameters as $\eta \sim \Delta P/\dot{\gamma}_{\infty}$ and the elastic modulus as $G \sim \Delta P/\gamma_0$. Further details about the fitting procedure can be found in our previous article \citep{Layachi2022}.  
The resulting fits, shown with dashed lines in the figure, display good agreement with the experimental data. 
The viscosity values range within  $4 \times 10^3$ and $4 \times 10^6$ Pa$\cdot$s (the mean and standard deviation are shown in panel (b)). The corresponding elastic moduli range within 50 and 400 Pa and the relaxation times within 0.5 and 2 s.  
The viscosity values obtained for SA-biotin prototissue are about 4000 larger than those for DNA-mediated prototissues. 
This large difference can be partly attributed, first, to the larger adhesion strength of an individual SA-biotin pair compared to the hybridization energy of the two complementary DNA strands (the ratio of the binding energies is approximately 10 times). 
One should also note that at this temperature ($\sim 25 ^{\circ}$C) DNA links are dynamic and they can spontaneously break and reform allowing for new GUV-GUV rearrangements. 
Thus DNA-mediated prototissues offer a very low resistance to flow.    
And second, the surface density of biotin molecules on GUV membranes is about 60 times larger than the surface functionalization of DNA strands, which thus allows for a larger number of bonds to be formed between the membranes of adjacent GUVs. 

Our results show that different assembly protocols lead to comparable rheological behavior of the synthetic prototissues but the values of the rheological parameters obtained for the prototissues may differ.  
The viscosity values reported in our previous work for in-bulk assembled prototissues, obtained with SA-biotin binding, were comprised within $3 \times 10^2$ and $10^5$ Pa$\cdot$s; the elastic modulus within $2$ and $2 \times 10^2$ Pa; and the relaxation time within 0.1 and 1 s. 
The rheological moduli reported in the present work for prototissues assembled in microfluidic confinement match the upper range of the rheological values obtained for the prototissues assembled in bulk. 
Larger values could be attributed to larger degrees of internal cohesion of the prototissues. In the present work we were able to optimize the bridging mechanism between GUVs to form cohesive prototissues, by collecting and maintaining GUVs in close contact inside the microfluidic trap during the introduction of the adhesion molecules. This facilitated the formation of adhesion patches between adjacent GUVs and of more cohesive structures.   
In addition, by achieving the assembly of the prototissue in the microfluidic device we removed any risk of rupture or internal damage of the prototissues that could be caused by shear flow during transfer or handling of the prototissues in the previous bulk method.  
The values of viscosities obtained for highly-adhesive prototissues are comparable to the range of viscosity values reported in the literature for cellular aggregates, which lies within $5 \times 10^3$ Pa s and $10^5$ Pa$\cdot$s \cite{Guevorkian2010,Kashkooli2021,Tlili2022}. In cellular aggregates, however, the role of cell cortex or active mechano-transduction mechanism could contribute  significantly in their rheological response \cite{Marmottant2009,Caorsi2016,Matejcic2023}.   

We can establish a correlation between the flow behavior observed at both global and local scales. 
DNA-mediated prototissues, for which we reported a significant degree of GUV-GUV reorganization at the local scale, display a linear increase of the deformation in time at the global scale.
Instead, SA-biotin mediated protissues, showed on the one hand significant deformation of individual GUVs at the local scale which would account for the elastic exponential response observed for the overall prototissue at short time scales. 
On the other hand, GUV-GUV reorganizations measured locally would account for the viscous behavior observed for the prototissue at longer time scales.    

Finally, the comparison of the overall shape of the prototissue before and after the microfluidic aspiration experiment informs us about the reversibility of the deformation. 
In Fig.\ \ref{fig:global}-c-d we provide a representative example of the contour of two prototissues obtained before/after passage of the microfluidic constriction, using DNA strands (left) or SA-biotin binding (right) for GUV-GUV assembly. The type of ligand used has a significant impact on the shape contour, as shown in the figure.  
For DNA-mediated prototissues, the low adhesiveness of the GUV-GUV interfaces together with the flow symmetry over an aspiration cycle permit that the prototissue maintains its overall integrity and revers back to a shape similar to the initial one. 
The width of the prototissue after the aspiration experiment (in the transverse direction of the flow) is (210  $\pm$ 30) $\mu$m, which corresponds to a reduction of approximately 27\% of its initial size. 
On the contrary, the SA-biotin prototissue undergoes irreversible (plastic) deformation. 
At high deformations GUV-GUV patches rupture in an irreversible way, most probably mediated by lipid pull-out from the GUV membranes, and can no longer reform. This leads to an elongated shape of the prototissue compared to its initial squared one. The typical width of the prototissue (in the transverse direction of the flow) after aspiration is (110  $\pm$ 40) $\mu$m, which is comparable to the width of the microfluidic constriction of the channel used for the experiments ($W_c = 100$ $\mu$m), and which corresponds to a reduction of approximately 57 \% of its initial size.  

\section*{Conclusions}
We have developed a microfluidic platform that can be successfully used to assemble artificial biomimetic tissues with tunable mechanical properties, and perform their rheological characterization simultaneously. 
Prototissues are obtained by the controlled assembly of GUVs. 
The utilization of the microfluidic device enables to assemble the GUVs in a sequential way: first by gathering an ensemble of GUVs in a microfluidic trap, which sets the size and shape of the prototissue, and next by triggering GUV-GUV adhesion by the incorporation of adhesion molecules.
The adhesion between GUVs have been mediated by the inclusion of DNA complementary strands or SA-biotin binding which provides weak and strong adhesion, respectively. 
The utilization of DNA or SA-biotin also have an impact on the dynamics of GUV-GUV binding. 
The dynamic nature of DNA-mediated adhesion provides GUV-GUV adhesion patches that can reform after eventual rupture under applied stress. On the contrary, SA-biotin mediated adhesion is irreversible and new bridging between GUVs after rupture is unfavorable.   
These differences at the molecular level, translate into qualitatively different rheological responses both at the scale of individual vesicles and of the entire prototissues, which can range from purely viscous to viscoelastic. 
On the one hand, we have shown that weakly-adhesive prototissues display at the local scale a response dominated by the reorganization of GUVs accompanied with very little deformation of individual GUVs. 
Local structure reorganizations allow to relieve stresses locally within the prototissue.   
This correlates with the behavior observed at the global scale which is also dominated by viscous dissipation and for which we have reported viscosity values not larger than 150 Pa$\cdot$s.   
On the other hand, highly-adhesive prototissues show at the local scale a viscoelastic response which results from a combination of GUV-GUV reorganizations and GUV elastic deformations. The adhesive contacts between GUVs slow down the dynamics of GUV reorganizations, which are less important in this case. While such reorganizations allow for dissipation of inner stresses, GUV deformations store  elastic energy within the prototissue.     
The viscoelastic behavior has also been recovered at the global scale, which encompasses a viscous and an elastically-dominated regimes at long and short timescales, respectively. Typical viscosity values for highly-adhesive prototissues range up to 4 $\times 10^6$ Pa$\cdot$s and elastic modulus up to 400 Pa. 
The rate of GUV-GUV rearrangements (and in turn the values obtained for the viscosity) could be affected by the polydispersity of GUVs constituting the prototissues, as it has been shown for flowing emulsions and granular systems \cite{Golovkova2021,Jiang2023}.  Although we have shown that the microfluidic platform allows us to diminish the polydispersity of GUVs, as compared to the polydispersity provided by a bulk assembly method, this could be further improved by synthesizing GUVs with encapsulation or microfluidic-based techniques \cite{Pautot2003,Abkarian2011,Matosevic2011,Cauter2021}. 
We have focused in this work on the role of GUV-GUV adhesion on the flow of GUV-prototissues. However, the elastic properties of GUV membranes could also have an impact on the flow behavior described.  
In the first place, the bending and stretching modulus of membranes (which depend on their lipidic composition \cite{Dimova2022}) determine the deformability of GUVs under flow. By changing the lipidic content we could thus also modify the flow response of the system.     
And even more importantly, GUVs are solely constituted of a lipidic bilayer and are devoid of an inner structuring network, akin to cell cytoskeleton. Artificial prototypes of cytoskeleton, based on actin filaments \cite{Pontani2009}, microtubule nematic gels \cite{Keber2014} or all-DNA gels \cite{Arulkumaran2023} have been successfully encapsulated into GUVs.  
The incorporation of an artificial skeletal framework would increase the stiffness of individual GUVs as well as their mechanical stability, and could also have a significant impact on the mechanical response of the prototissue at the global scale.  

Overall, the microfluidic platform that we have developed enables to synthesize biomimetic artificial prototissues and to perform their mechanical characterization in the same device. 
Biomimetic prototissues present a reduced complexity but offer the possibility to fine-tune the strength of the adhesion between cells and to tune the size and shape of the tissues. Besides, prototissues can be designed to display a purely viscous flow behavior or a viscoelastic response, comprising both elastic and viscous contributions, comparable to those reported for cell aggregates.  
Given the versatility offered by this approach, we believe such a microfluidic platform combined to model vesicle prototissues constitute a powerful toolbox to rationalize the mechanics of cell aggregates, and eventually of living tissues, in controlled flow settings.    

\section*{Author contributions}
LC led the project. ML performed experiments and data analysis. RM designed the DNA sequences used in the experiments. 
ML and LC wrote the first draft of the manuscript and all the authors reviewed and edited the manuscript. 
All authors contributed to scientific discussions of the results. 

\section*{Conflicts of interest}
There are no conflicts of interest.

\section*{Data availability} 
The data supporting this study is included within the article and the ESI\dag. The scripts used for data analysis are included in a GitHub deposit (https$://$gitlab.com/mlay0132/script-Deposit$\_$art-L2C-202507.git). 
Further information (including raw microscopy data) can be made available upon request from the authors.

\section*{Acknowledgments}
We acknowledge the researchers from the Soft Matter team at L2C for fruitful discussion and C. Blanc for the help provided on the microfabrication. We thank F. Graner for a critical reading of the manuscript. 
This project was supported by the \textit{Agence National de la Recherche} (grant ANR-21-CE06-0023–01). 

%%%END OF MAIN TEXT%%%

\balance

%%%REFERENCES%%%
\bibliography{BiblioLC_vfinal_noURL} 
\bibliographystyle{unsrtnat} %abbrv %ieeetr

% --- Supplementary material starts here ---
\clearpage  % start on a new page
\onecolumn   % optional: switch to one column for supplementary

\section*{Supplementary Material}

\subsection*{Methods: \textit{Angularity} shape descriptor}
\begin{figure}[h!]
\centering 
	\includegraphics[height=18cm]{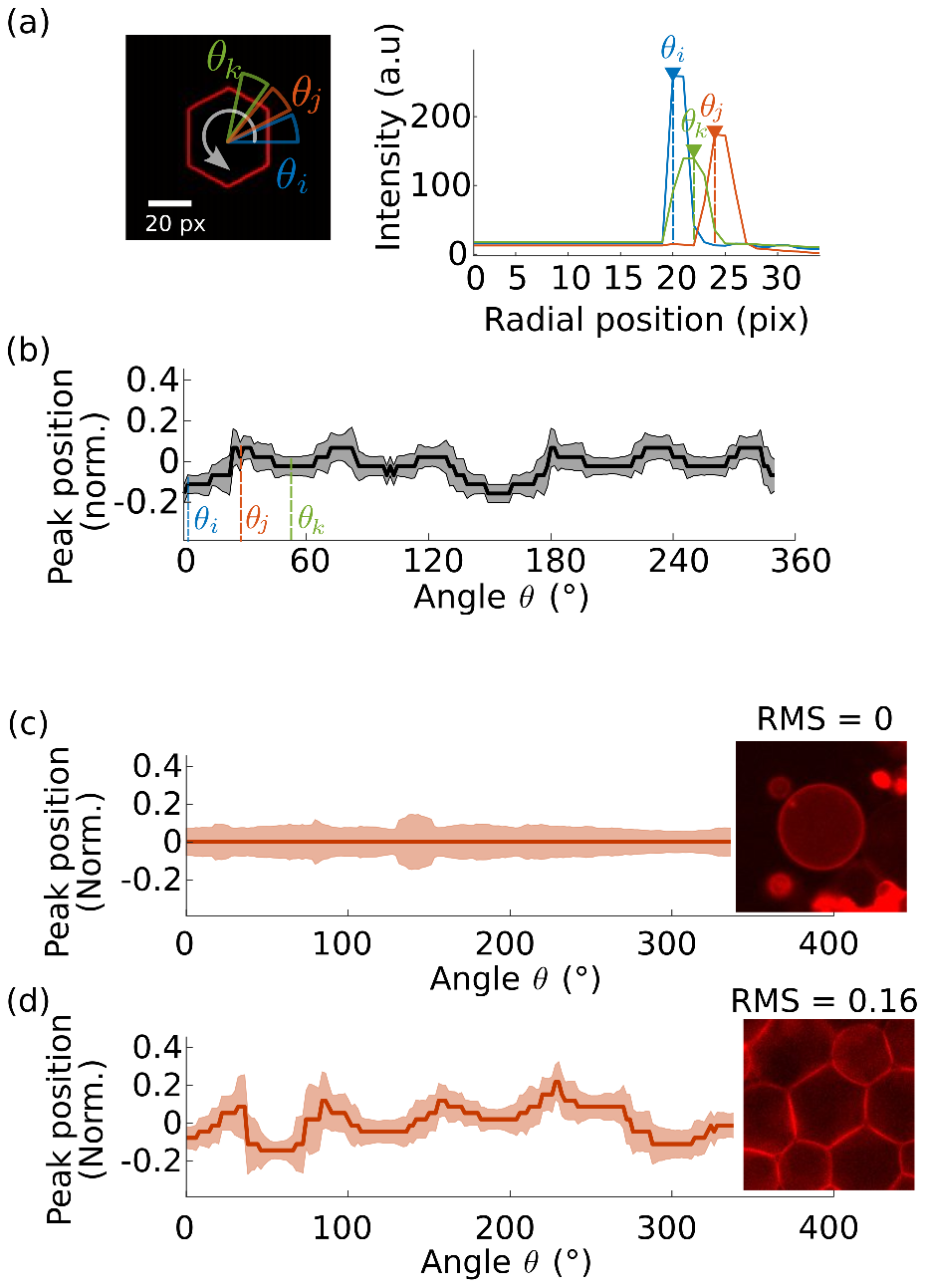}
	\caption{(a). An example of the computation of the angularity ($\Phi$) is shown for an object with an hexagonal contour. The contour is displayed in red. The intensity of the object contour is averaged over an arch of circumference of 10$^{\circ}$. The entire contour is covered using the necessary number of arches with an overlap of {8}$^{\circ}$. Three different arches are shown in the left of the panel (named with \textit{i,j,k} subscripts) as an example. The corresponding intensity profiles are plotted as a function of the radial coordinate (in pixels) and the radial location corresponding to the maximum intensity is identified for each arch. 
(b). The position of the intensity peak (in dimensionless units) is plotted as a function of the angle, spanning over an entire revolution from $0^{\circ}$ to $360^{\circ}$. The $RMS$ of the peak position is used as the shape descriptor. (c-d). Peak position obtained for the membrane of two different GUVs as a function of the angle. The corresponding RMS values are shown. The two examples correspond to an individual GUV displaying an almost spherical contour (c), and a GUV displaying a polygonal shape (d) which is contained within a prototissue (the analyzed GUV is enclosed in the yellow circle).}  
  \label{fig:SI_MM}
	\end{figure}

We define a shape descriptor, that we name \textit{angularity} ($\Phi$), which enables to distinguish circular (smooth) GUV contours from polygonal (angular) ones. 
In Fig.\ \ref{fig:SI_MM} we describe the methodology used to compute $\Phi$ as well as some examples obtained for different GUV contours. In brief, this descriptor quantifies the departure of a vesicle contour from a circular shape, by computing the number of bright pixels on the contour of the object lying away from an ideal circle. 

\subsection*{Methods: DNA sequences} 
In Tab.\ \ref{tab:seqsDNA} we include the sequences of the DNA strands used: the two sequences A1 and A2 constituting the anchoring construct; and the  sequence (B) corresponding to the linker, which contains the palindromic part (highlighted in bold letters). 
In Tab.\ \ref{tab:energyDNA} we display the values for the total free energy $\Delta G$ (including both enthalpic ($\Delta H$) and entropic ($\Delta S$) terms) corresponding to the palindromic part of the linker and the anchoring construct. These values have been calculated with the Unafold software at a DNA concentration of 1 mM and 150 mM of NaCl, at 25 $^{\circ}$ C. As shown in the table, the free energy corresponding to the palindromic sequence is much lower than the energy corresponding to the anchoring construct.   

\begin{landscape}
%[insert table here that will be displayed horizontally]
\begin{table}[h!]
\begin{center}
\begin{tabular}[h] {ccccc}
%\hline
Name & Sequence		($5'\rightarrow 3'$)& 	$5'$ modification & $3'$ modification & Purification\\
%\hline
A1      &	  \textbf{CGATATCG}AACCGACAGTGATTCCATGCACGA  & Atto488 & - & RP-HPLC\\  			
A2      &	  ATTCACTAGAAGTTGAAGACTTACTA\textit{TTTT} &  Atto565 & Cholesteryl-TEG & RP-HPLC  \\ 
B       &	  \textit{TTTT}TAGTAAGTCTTCAACTTCTAGTGAATTCGTGCATGGAATCACTGTCGGTT & Cholesteryl-TEG  & - & RP-HPLC\\ 
%\hline
\end{tabular}
\caption{DNA sequences used to design the constructs for GUV-GUV binding. The part of the sequence in italic letters is a single-stranded region which ensures flexibility at the anchoring site with the  bilayer. The palindromic part is highlighted in bold letters.}
\label{tab:seqsDNA}
\end{center}
\end{table}

\vspace{2cm}

\begin{table}[h!]
\begin{center}
\begin{tabular}[h] {cccccc}
%\hline
Sequence &	Tm ($^{\circ}C$) 	& $\Delta H$ (kcal/mol)	& $\Delta S$ (cal/(mol $\cdot$ K))	& $\Delta G$ (kcal/mol) & $K_BT$ units\\
CGATATCG	& 41 &	31.1	& 98.5	& 1.8	& 3.0 \\
AACCGACAGTGATTCCATGCACGA &	80.2 &	202.3 &	576.8 &	30.4	& 51.6  \\
ATTCACTAGAAGTTGAAGACTTACTA &	72.9 &	207.5 &	607.7 &	26.41 &	44.8 \\ 		
\end{tabular}
\caption{Free energy terms computed for the palindromic sequence and the anchoring constructs.} 
\label{tab:energyDNA}
\end{center}
\end{table}
\end{landscape}

\end{document}